\documentclass[sigconf]{acmart}

\AtBeginDocument{%
  \providecommand\BibTeX{{%
    Bib\TeX}}}

\setcopyright{acmcopyright}
\copyrightyear{2023}
\acmYear{2023}
\acmDOI{XXXXXXX.XXXXXXX}

\acmConference[CIKM '23]{32nd ACM International Conference on Information and Knowledge Management }{October 21--25,
  2023}{Birmingham, UK}
\acmPrice{15.00}
\acmISBN{978-1-4503-XXXX-X/18/06}

\usepackage{cleveref}
\usepackage{xcolor,listings}
\begin{document}

\title{SieveJoin: Boosting Multi-Way Joins with Reusable Bloom Filters} 

 \author{Qingzhi Ma}
 \affiliation{%
  \institution{Soochow University}
   \city{Soochow}
    \country{China}
 }

 \email{qzma@suda.edu.cn}

\renewcommand{\shortauthors}{}

\begin{abstract}
Improving data systems' performance for join operations has long been an issue of great importance.
More recently, a lot of focus has been devoted to multi-way join performance and especially on reducing the negative impact of
producing intermediate tuples, which in the end do not make it in the final result.
We contribute a new multi-way join algorithm, coined SieveJoin, which extends the well-known Bloomjoin algorithm to multi-way joins and achieves state-of-the-art performance in terms of join query execution efficiency.
SieveJoin's salient novel feature is that it allows the propagation of Bloom filters in the join path, enabling the system to `stop early' and eliminate useless intermediate join results.
The key design objective of SieveJoin is to efficiently `learn' the join results, based on Bloom filters, with negligible memory overheads. 
We discuss the bottlenecks in delaying multi-way joins, and how Bloom filters are used to remove the generation of unnecessary intermediate join results. 
We provide a detailed experimental evaluation using various datasets, against a state-of-the-art column-store database and a multi-way worst-case optimal join algorithm, showcasing SieveJoin's gains in terms of response time.  
\end{abstract}




\maketitle
\section{Introduction}
\label{sec:intro}
Queries involving multiple tables are being intensively used for data mining and decision-making. The majority of database systems purely rely on binary join plans to process multi-way joins \cite{nes2012monetdb,stonebraker1986design,lahiri2013oracle}. Binary join plans execute the join operation on two relations at a time. They 
have been extensively studied and optimized over the past several decades \cite{idris2017dynamic,yannakakis1981algorithms,goodman1983gyo}. As a consequence,  they provide great flexibility, high reliability, and robustness for a wide range of scenarios. Alas,  binary joins are sub-optimal from a theoretical point of view \cite{ngo2018worst,ngo2014skew}, and such a phenomenon is being continuously discovered and discussed in recent studies \cite{graefe1998hash,veldhuizen2014leapfrog,freitag2020adopting,shanghooshabad2022graphical}. Specifically, binary joins suffer from generating an exploding number of intermediate tuples which never make it to the final join result, 
leading to large wastes of computational resources and much higher query response times \cite{freitag2020adopting}.

In contrast, more recent research on multi-way joins aims to avoid the generation of useless intermediate join results by accessing multiple tables at the same time  \cite{avnur2000eddies, aberger2017emptyheaded, aberger2018levelheaded}. 
The top efforts have been applied (or considered ) to commercial database systems \cite{graefe1998hash, kemper1999generalised, freitag2020adopting,veldhuizen2014leapfrog, zukowski2012vectorwise}, but some are dropped due to system  stability considerations and performance issues, like Microsoft SQL Server \cite{kemper1999generalised}.  Recently, theoretical progress in database systems gave rise to worst-case optimal join algorithms (WCOJ), which exhibit better asymptotic runtime complexity compared to binary joins \cite{ngo2018worst,ngo2014skew,ngo2014beyond}. Such efforts  include the well-known leapfrog algorithm \cite{veldhuizen2014leapfrog,chu2015theory,wu2014multipredicate}, the Umbra database system \cite{freitag2020adopting}, etc\cite{shanghooshabad2022graphical,zou2011gstore}. Such methods have great potential to outperform binary joins by  filtering intermediate join results. 
However, the vast majority of these efforts show non-negligible disadvantages, which greatly limit their adoption into database systems. 
For instance, WCOJ performs even worse for cases without many intermediate results.
Also, additional index structures are usually used, which causes performance issues for workloads with frequent updates. 

Another category of efforts to boost join processing involves the usage of Bloom filters. A Bloom filter \cite{bloom1970space,kirsch2006less,luo2018optimizing} is a space-efficient data structure for fast membership checks. 
In the context of table joins, if join results could be pruned in advance so that useless intermediate join results are filtered out, the total query processing time will be reduced significantly. 
Bloomjoin \cite{bratbergsengen1984hashing,mackert1986r} was the early study that adopted Bloom filters to reduce data transferred between sites. But its usage was limited to two-relation joins.
\cite{mackert1986r} shows that Bloomjoin consistently outperforms the basic semi-join algorithm. 
Bloomjoin was discovered to save costs where there is a significant amount of tuples to be filtered, which is not always the case \cite{mullin1990optimal}.
Thus, an incremental Bloom filter was proposed to reduce the transmission cost.
Bloom filter operations \cite{koutris2011bloom, michael2007improving} are used for improving multi-way join performance in the distributed environment, and they introduced new bloom filter operations to support this. Specifically, 
bitwise AND is used to compute the intersections requires by joins, and bitwise OR is used to compute unions in case of horizontal fragmentation. However, such operations could only be applied to the same join attribute.  In addition, they did not provide solid implementation and experiment results to support the idea. 
LIP \cite{zhu2017looking} relies on Bloom filters to generate a robust and efficient query plan for multi-way joins, the work focuses on the order of applying Bloom filters and is limited to  star schema. 
Bitvector filtering \cite{ding2020bitvector} further analyzes a much broader range
of decision support queries (star and snowflake queries) and plan search space, and reduces CPU execution time significantly.
These query optimization techniques rely on binary join plans. Also, no support for cyclic joins is provided, or discussed.

Bloom filters are also applied to database systems to augment other functionalities.
Spectral Bloom filters \cite{cohen2003spectral} are used for fast membership queries, and are not used for join processing.
Bloom filter indexes are also applied in Databricks \cite{etaati2019azure}, Apache Doris, Postgres \cite{stonebraker1986design}, etc. Again, they are used for fast membership queries over multiple columns.
Vectorwise uses bloom filters to accelerate hash-table lookups in situations where the key is often not found \cite{ruaducanu2013micro}.
And the discussion or extension to multi-way joins is not mentioned.

\section{The SieveJoin Algorithm}

\subsection{The Bloomjoin Algorithm}
\label{sec:bloomjoin:intro}
\begin{figure}[ht]
    \centering
    \includegraphics[width=0.48\linewidth]{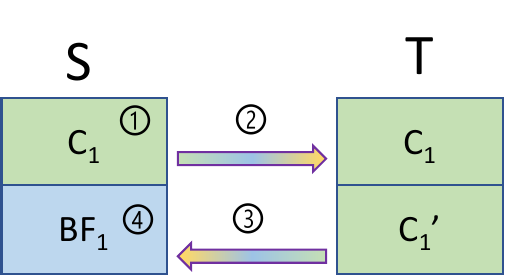}
    \vspace{-0.1cm}
    \caption{The Bloomjoin Algorithm}
    \label{fig:bloomjoin}
\end{figure}

In this section, we will recall the well-known Bloomjoin algorithm\cite{mackert1986r,bratbergsengen1984hashing}, which inspired our work here.
It is designed for two-way joins.
Assuming there are two relations S and T, with the join condition on $S.C_1=T.C_1$. 
\Cref{fig:bloomjoin} shows the typical Bloomjoin algorithm, which consists of four steps.

\vspace{-0.3cm}

\begin{enumerate}
    \item \textbf{Generate a Bloom filter $BF_1$ on attribute $C_1$ within relation S.} The Bloom filter is a large vector of bits, which is initially set to `0'. By the linear scanning and hashing of every tuple in Attribute $C_1$, the corresponding bit within $BF_1$ is set to `1'.
    \item \textbf{Send $BF_1$ to Relation T.} The size of the Bloom filter is 1 bit per tuple, which is significantly smaller than transferring the actual relation S.
    \item \textbf{Hasing every tuple $t_i$ of Attribute $C_1$ in Relation T. If $BF_1$.contains($t_i)$, send this tuple to Relation S as tuple stream $C_1{'}$.}
    The expected number of tuples in $C_1{'}$, or the Bloomjoin cardinality $BC$ of Relation $C_1{'}$, is given by
    \begin{equation}
        BC=SC_T + bits_S(1-e^{-\frac{\alpha D_T}{F}})
    \end{equation}
    where 
    \begin{equation}
        \alpha = 1-\frac{SC_T}{C_T}
    \end{equation}
    is the fraction of non-matching tuples in T,  
    \begin{equation}
        bits_S = F(1-e^{-\frac{D_S}{F}})
    \end{equation}
    is the approximate number of bits set to `1' in $BF_1$ \cite{severance1976differential},  $SC_T$ is the semijoin cardinality of T, F is the size of $BF_1$, $D_T$ is the number of distinct values of Attribute $C_i$ in Relation T.
    \item \textbf{Join the tuple stream $C_1{'}$ to S, and return the join result to the user.}
\end{enumerate}
 Suppose there are indexes on $S.C_1$ and $T.C_1$. In that case, we could enhance the Bloomjoin algorithm as mentioned above by accessing the indexes instead, thus saving the cost of accessing data pages in Step (1),  and boosting the join operation in Step (4).

 The Bloomjoin algorithm is designed for two-table distributed equi-joins, and is exceptionally advantageous for cases where the message costs are high and any table remote from the join site is quite large. 
 One could envisage a straightforward application of Bloomjoin for n-way joins, as follows: Produce a binary-join plan for the n-way join and repeat the algorithm for each two-way join in the plan. However, this is still a binary-join plan, suffering from all aforementioned drawbacks. 
 
\subsection{SieveJoin}
Motivated by the Bloomjoin algorithm as mentioned above, we propose a `learning and pruning' 
SieveJoin algorithm that extends the application of Bloom filters to multi-way joins.
To simplify the problem, we will re-use relations $R, S$ and $T$, as introduced in \Cref{sec:bloomjoin:intro}.
Our target is to boost the query processing of the multiple join $Q:=R\bowtie_b S\bowtie_c T$.
\begin{figure}[h]
    \vspace{-0.2cm}
    \centering
    \includegraphics[width=0.7\linewidth]{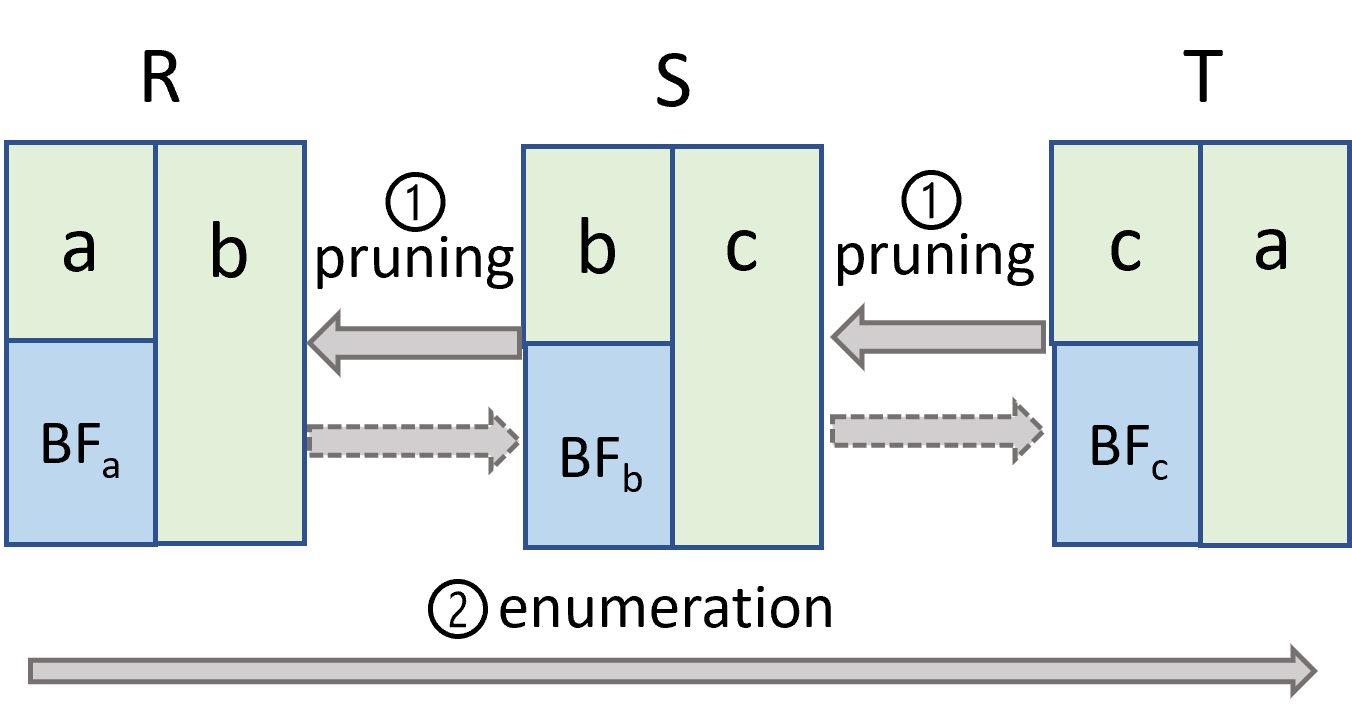}
    \caption{SieveJoin's training flow}
    \label{fig:training_flow}
    \vspace{-0.2cm}
    
\end{figure}

\Cref{fig:training_flow} shows the query execution flow of SieveJoin. 
There are two phases in SieveJoin: the  pruning and enumeration phases.
\begin{enumerate}
    
    \item \textbf{Pruning Phase}. In this phase, Bloom filters are generated and pruned backward (from the right-most relation to the left-most relation).
    Specifically, we generate a Bloom filter $BF_c$ on $T.c$. After that,  Relation $S$ receives $BF_c$,  which is used to  generate and prune the Bloom filter $BF_b$ on $S.b$. Note, $BF_b$ now holds information of the sub join query $S\bowtie_cT$ instead of Relation $S$ only. In other words, not all tuples in $S$ are used to generate $BF_b$, tuples that belong to $S$ and occur in $S\bowtie_cT$ are used. Thus, Bloom filter $BF_b$ is pruned to remove unneeded intermediate join results.  We repeat this pruning phase from  $S$ to $R$ to generate Bloom filter $BF_a$ on $R.a$.  

    \item \textbf{Enumeration Phase}. In this phase, the pre-generated Bloom filters are used to avoid the generation of unneeded intermediate results and produce the final join results. In this study, SieveJoin is embarrassingly implemented based on the binary join plan.
\end{enumerate}


The pruning phase relies on pre-built Bloom filters to prune the current relation and/or Bloom filter. A bit-wise ADD operation might be used in this stage for faster training. However, this is not an ideal solution as the cardinalities of attributes within/across relations vary, yielding Bloom filters of different sizes.
SieveJoin generates the pruned Bloom filters by a linear scan of the target attribute,  in the meantime, accessing other relevant Bloom filters as necessary.

Note, SieveJoin adopts a simple one-pass backward pruning strategy. Another forward pruning phase (from left-most to right-most, refer to the dotted arrow in \Cref{fig:training_flow}) might be applied to further prune relations on the right side, at the cost of doubling the training time and  space overheads of Bloom filters for additional columns. This two-pass pruning strategy is useful for distributed environments where bandwidth is limited.

\begin{figure*}[ht]
    \centering
    \includegraphics[width=0.9\linewidth]{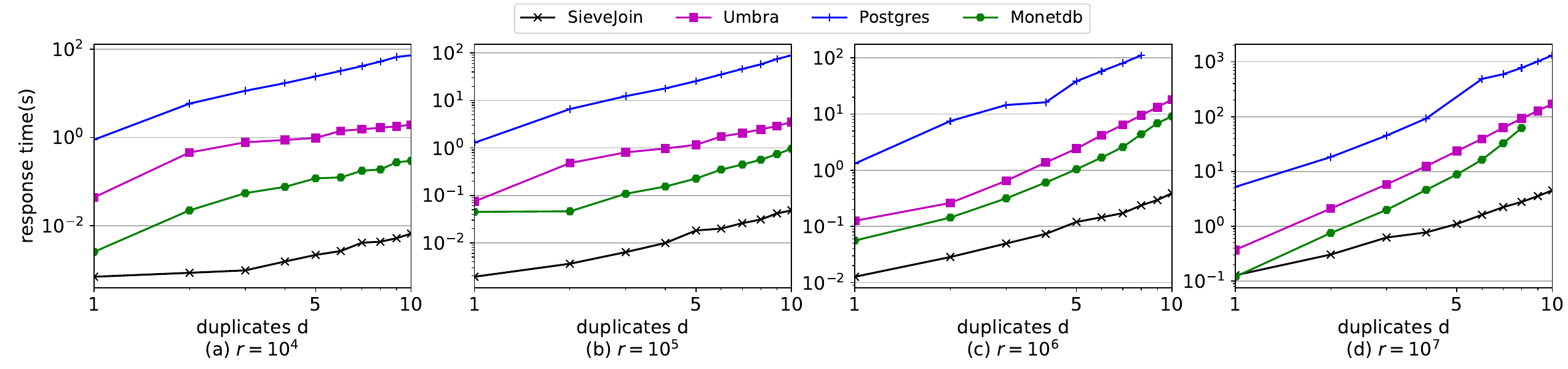}
    \vspace{-0.2cm}
    \caption{Comparison of Query Response Time for the Synthetic Data}
    \label{fig:rst_response}
\end{figure*}

Suppose there are indexes built on join attributes, we could boost the pruning phase by accessing the indexes directly. 
SieveJoin is applicable to existing join enumeration algorithms, including nested loop join, index-based nested loop join \cite{patel1996partition, dewitt1993nested}, hash join \cite{lo1996spatial,blanas2011design,barber2014memory}, other multi-way joins \cite{freitag2020adopting, graefe1998hash}, etc. 

\subsection{Reuse of Bloom Filters}

\begin{figure}[ht]
    \centering
    \includegraphics[width=0.9\linewidth]{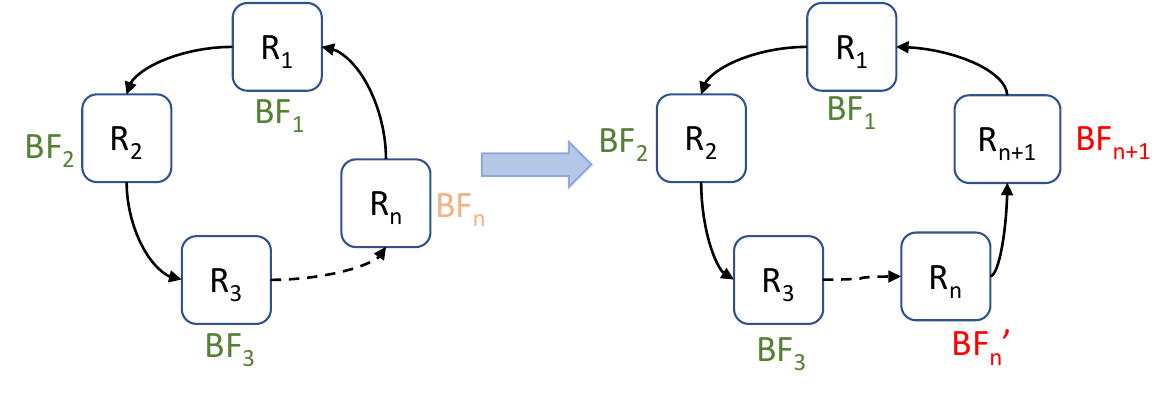}
    \vspace{-0.4cm}
    \caption{Reuse of Bloom filters for cyclic joins}
    \label{fig:cycle}
    \vspace{-0.5cm}
\end{figure}

SieveJoin is designed for popular queries with predictable workloads, and might also be applied to solve ad hoc queries, where applicable.
Specifically, assuming we have generated Bloom filters BFs for a popular query template $Q_1$. These pre-generated BFs could be reused for an ad hoc query $Q_2$, if $Q_1$ is a sub-query of $Q_2$.
The application to acyclic queries is achieved straightforwardly.
For cyclic queries, 
we could only reuse part of the pre-generated BFs, while the Bloom filter belonging to the final closing condition has to be re-generated on the fly.
\Cref{fig:cycle} demonstrates the reuse strategy for cyclic joins in more detail.
We  reuse the majority of the pre-generated BFs ($BF_1$ to $BF_{n-1}$), and pay extra time to generate Bloom filters for $R_n$ and $R_{n+1}$ on the fly.

\subsection{Support for Updates}
\label{sec:update}
As Bloom filters are the only extra data structures to existing join algorithms, the performance of handling updates is greatly influenced by Bloom filters. 
Luckily, Bloom filters are highly efficient data structures and the time complexity to insert a new tuple is $O(1)$.
Assuming a new tuple $t_1$ is to be inserted to Relation $R_i$, which holds the pre-generated Bloom filter $BF_1$.
We just need to check if $BF_1.contains(t_1)$ holds. 
If true,  we do not need to update the Bloom filter. And the complexity is $O(1)$.
Otherwise, we have to search from Relation $R_i$ to its adjacent relations, with complexity   $O(n)$, where $n$ is the number of relations to be joined.

Currently, SieveJoin based on traditional Bloom filters only supports frequent insertions. 
If counting Bloom filters, which hold a counter for each existing tuple, are applied to SieveJoin, deletions and updates will be supported straightforwardly, with the cost of slightly larger space overheads.

\section{Performance Evaluation}

\subsection{Experimental Setup}
In the following, we present a comprehensive evaluation of the implementation of the proposed SieveJoin algorithm. 
SieveJoin \footnote{code is available at \url{https://anonymous.4open.science/r/SieveJoin-314B}} is implemented in C++. It adopts the column-oriented fashion where columns are stored in a list of vectors. It is an in-memory demo database without a query parser. Thus, we only report the query execution time in this study.
All experiments are run on a Ubuntu 20.04 server  with 16 CPU cores (32 threads) on an AMD 7950X processor, 128GiB RAM, and 4TB of SSD disk space. We repeat each experiment three times and report the average values. Unless specified, all systems run in parallel.

\subsubsection*{\textbf{Baselines:}}\hfill\\

\begin{itemize}
    \vspace{-0.5cm}
    \item Umbra \cite{freitag2020adopting}: the state-of-the-art multi-way join database, which uses a heuristic optimizer to trade-off between WCOJ  and binary join plans;
    \item MonetDB \cite{boncz2005monetdb,nes2012monetdb}: the well-known column-store database, which exclusively relies on binary join plans.
    \item Postgres \cite{stonebraker1986design}: a free and open-source relational database management system (RDBMS) based on  binary join plans.
\end{itemize}

For the experiment, we first evaluate SieveJoin against other database systems over a well-designed synthetic dataset \cite{freitag2020adopting}, which controls the number of intermediate join results exactly. And the size of all relations is about 40GiB.
Then we  run experiments on  the well-known TPC-H dataset \cite{poess2000new}.
Experimental results over other datasets, like the popular Wikipedia vote network \cite{leskovec2010signed, leskovec2010predicting} and the citation network \cite{leskovec2005graphs,gehrke2003overview}, produce consistent results, and we apply a trick to further improve SieveJoin's performance. Some of the experimental results are omitted for space reasons.



\subsubsection{Synthetic Workload}\hfill\\
To demonstrate the performance of SieveJoin for cases with a controlled number of intermediate join results, we first conduct experiments on the synthetic benchmark as described in \cite{freitag2020adopting}.
Here we  give a brief introduction to this synthetic benchmark.
Two parameters, $r\in \{10^4, 10^5, 10^6, 10^7\}$,  and $d \in \{1, ... , 10\}$ are used to control the number of intermediate join results. 
Three relations $R$, $S$, and $T$ are generated as follows.
$R$ simply contains the integers from 1 to $10^7$, $S$ contains integers from 1 to $(10^7+r)/2$, and $T$ contains integers from $(10^7-r)/2$ to $10^7$, respectively.
Each integer in $R$, $S$, and $T$ is duplicated $d$ times.
Here, \textit{$r$ controls the scale of this dataset, while $d$ controls the complexity/duplicates of this dataset.}
The resulting natural join $R \bowtie S \bowtie T$ contain exactly $r$ distinct integers, each integer is duplicated $d^3$ times. Thus the final join results contain $rd^3$ tuples.

\Cref{fig:rst_response} compares the  response time of the query $R \bowtie S \bowtie T$ between different database systems, including Postgres, MonetDB, SieveJoin, and state-of-the-art multi-way join database Umbra.
As $r$ increases from $10^4$ to $10^7$, the query response times from all systems increase correspondingly. For instance, the query response time of Postgres increases to more than 1000s for the case when $r=10^7$ and $d=10$.
$d$ controls the  number of duplicates.  Recall that the number of join results contains $rd^3$ tuples. With an increased number of $d$, the query response time from all systems increases.

Clearly, Postgres performs the worst, as it relies purely on binary join plans.
There are even some missing values for Postgres and MonetDB as the system runs out of memory.
Umbra is based on a multi-way join implementation that meets the worst-case optimal criteria, and is more robust with increased duplicates. 
Interestingly, the performance of Umbra is slightly worse than MonetDB.
For SieveJoin, its query response time grows much slower as it has the ability to `stop early' and avoid the generation of exploding intermediate join results.

Note, for popular query workloads, SieveJoin performs the best as Bloom filters are prepared in advance. SieveJoin might also boost the processing of  ad hoc queries where pre-generated  Bloom filters (for other queries) might be used. In other words, Bloom filters act like traditional index structures in database systems to boost tuple positioning and query processing, but for a multi-way join scenario. 


\subsubsection{TPC-H Workload}\hfill\\
\label{sec:tpch}
We used scale factor SF=1, resulting in the largest table having $\approx$ 6 Million rows. 
To control the proportion of intermediate join results, we make samples of size $\in \{100\%, 80\%, 60\%, 40\%, 20\%, 10\%, 1\%, 0.1\%     \}$ for table \texttt{lineitem}, which generates $\approx$ $\{0.6\%, 5\%, 10\%, 20\%, 40\%, 60\%$, $80\%, 95\% \}$ of intermediate join results.
We run acyclic joins of query X involving 5 tables, as described in \cite{zhao2018random}. Specifically, this query aims to find suppliers and customers in the same nations with the purchase history of the customers.
Note, we also run experiments for SF=10 and 40, but as the queries take a much longer time to execute, and sometimes cause timeouts, these results are discarded.


\begin{figure}[ht!]
    \centering
    \vspace{-0.4cm}
    \includegraphics[width=0.7\linewidth]{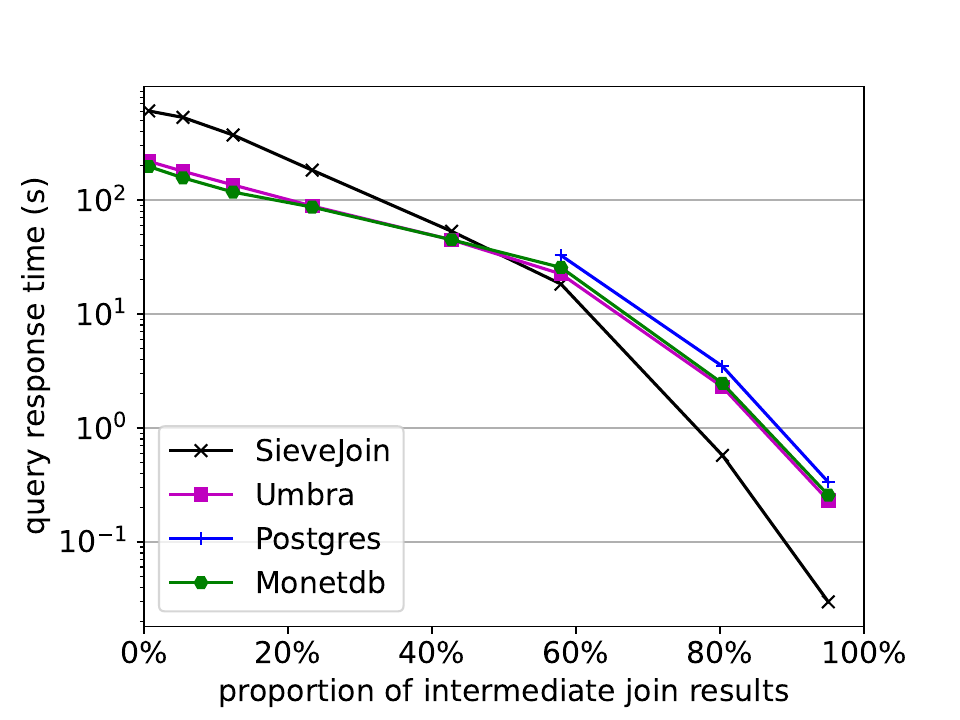}
    \vspace{-0.2cm}
    \caption{Comparison of Query Response Time for TPC-H}
    \vspace{-0.3cm}
    \label{fig:tpch}
\end{figure}
\Cref{fig:tpch} summarizes the query response time from different systems. This time, all systems run on a single core.
As the proportion of intermediate join results increases, query response time drops.
This is expected as fewer tuples are processed and generated during query processing, showcasing various database systems' abilities to avoid the generation of intermediate join results.
The performance of Umbra and MonetDB is similar. 
Postgres performs the worst, and the system runs out of memory (128GB) in many cases.
The picture for SieveJoin is quite different this time.
For cases with a large number of  intermediate results, SieveJoins achieves 1 order of magnitude savings in query response time, demonstrating SieveJoin's strong ability to avoid intermediate results.
For cases where there are few intermediate results, SieveJoin is much slower than other systems. This is embarrassingly an implementation issue, as SieveJoin is currently based on binary join plans. We will demonstrate how to overcome this issue in the next experiment.

\subsubsection{Wikipedia Vote Network}\hfill\\
The Wikipedia vote network \cite{leskovec2010signed, leskovec2010predicting} contains the voting records of users for adminship of the free encyclopedia.  Each row in the dataset contains two user ids, indicating user1 votes user2 as an administrator. This gave us 2,794 elections with 103,663 total votes and 7,066 users participating in the elections. In this experiment, we run the 3-clique and 4-clique queries.
\vspace{-0.4cm}
\begin{figure}[ht!]
    \centering
    \includegraphics[width=0.7\linewidth]{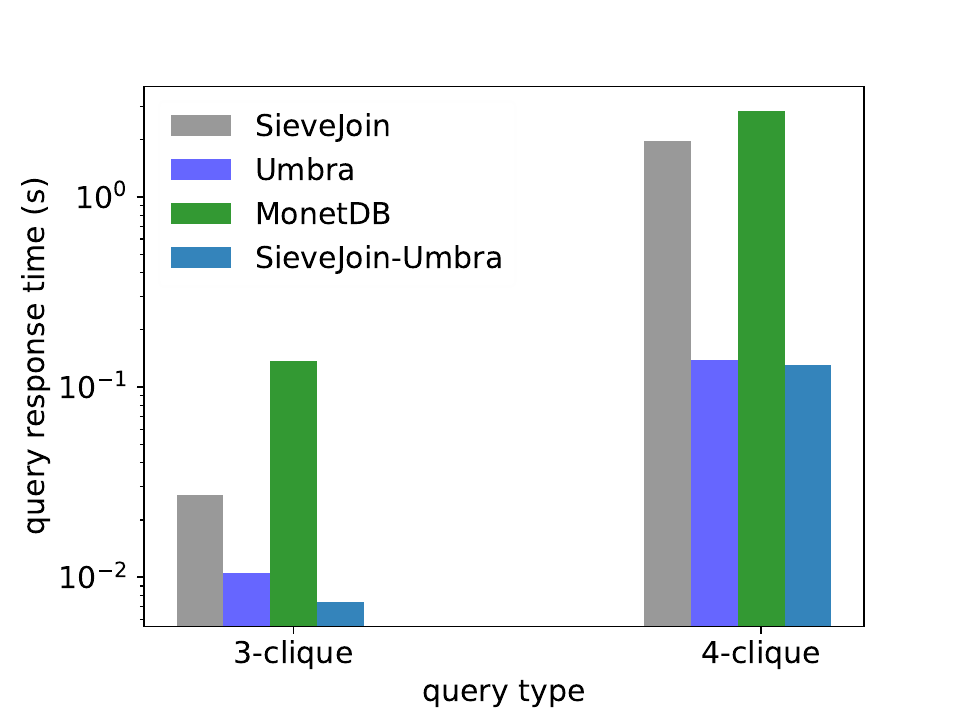}
    \vspace{-0.2cm}
    \caption{Comparison of Query Response Time for Wikipedia}
    \label{fig:wiki}
    \vspace{-0.3cm}
\end{figure}

As described earlier, SieveJoin relies on Bloom filters to avoid intermediate results, and is able to \textit{prune tables} effectively.
Motivated by this idea, we feed tables pruned by SieveJoin to Umbra,
as if SieveJoin was implemented based on a multi-way join plan (specifically, WCOJ), and this is denoted as SieveJoin-Umbra.
\Cref{fig:wiki} compares the query response time from various systems.
SieveJoin outperforms MonetDB but is worse than Umbra. 
As pointed out earlier, SieveJoin is currently implemented based on binary join plans.
If SieveJoin is implemented based on a multi-way join plan, as denoted SieveJoin-Umbra, it beats all other systems.

\vspace{-0.25cm}
\section{Conclusion}
With this paper, we presented SieveJoin, a Bloom filter based multiple join boosting algorithm.
SieveJoin's salient feature is that it relies on Bloom filters to `stop early' and eliminate useless intermediate join results.
The key insight is that 
the usage of pre-generated Bloom filters is not limited to a specific query, but could also be reused for other ad hoc queries, where applicable.
These facts allow SieveJoin to offer dramatic speedups, while being very frugal in memory requirements, as Bloom filters are compact.
Future work focuses on 
better integration of SieveJoin with other databases based on multi-way joins, and the optimization of the pruning phase of Bloom filters.



\balance
\bibliographystyle{ACM-Reference-Format}
\bibliography{main}


\begin{thebibliography}{47}


\ifx \showCODEN    \undefined \def \showCODEN     #1{\unskip}     \fi
\ifx \showDOI      \undefined \def \showDOI       #1{#1}\fi
\ifx \showISBNx    \undefined \def \showISBNx     #1{\unskip}     \fi
\ifx \showISBNxiii \undefined \def \showISBNxiii  #1{\unskip}     \fi
\ifx \showISSN     \undefined \def \showISSN      #1{\unskip}     \fi
\ifx \showLCCN     \undefined \def \showLCCN      #1{\unskip}     \fi
\ifx \shownote     \undefined \def \shownote      #1{#1}          \fi
\ifx \showarticletitle \undefined \def \showarticletitle #1{#1}   \fi
\ifx \showURL      \undefined \def \showURL       {\relax}        \fi
\providecommand\bibfield[2]{#2}
\providecommand\bibinfo[2]{#2}
\providecommand\natexlab[1]{#1}
\providecommand\showeprint[2][]{arXiv:#2}

\bibitem[Aberger et~al\mbox{.}(2018)]%
        {aberger2018levelheaded}
\bibfield{author}{\bibinfo{person}{Christopher Aberger},
  \bibinfo{person}{Andrew Lamb}, \bibinfo{person}{Kunle Olukotun}, {and}
  \bibinfo{person}{Christopher R{\'e}}.} \bibinfo{year}{2018}\natexlab{}.
\newblock \showarticletitle{Levelheaded: A unified engine for business
  intelligence and linear algebra querying}. In \bibinfo{booktitle}{\emph{2018
  IEEE 34th International Conference on Data Engineering (ICDE)}}. IEEE,
  \bibinfo{pages}{449--460}.
\newblock


\bibitem[Aberger et~al\mbox{.}(2017)]%
        {aberger2017emptyheaded}
\bibfield{author}{\bibinfo{person}{Christopher~R Aberger},
  \bibinfo{person}{Andrew Lamb}, \bibinfo{person}{Susan Tu},
  \bibinfo{person}{Andres N{\"o}tzli}, \bibinfo{person}{Kunle Olukotun}, {and}
  \bibinfo{person}{Christopher R{\'e}}.} \bibinfo{year}{2017}\natexlab{}.
\newblock \showarticletitle{Emptyheaded: A relational engine for graph
  processing}.
\newblock \bibinfo{journal}{\emph{ACM Transactions on Database Systems (TODS)}}
  \bibinfo{volume}{42}, \bibinfo{number}{4} (\bibinfo{year}{2017}),
  \bibinfo{pages}{1--44}.
\newblock


\bibitem[Avnur and Hellerstein(2000)]%
        {avnur2000eddies}
\bibfield{author}{\bibinfo{person}{Ron Avnur} {and} \bibinfo{person}{Joseph~M
  Hellerstein}.} \bibinfo{year}{2000}\natexlab{}.
\newblock \showarticletitle{Eddies: Continuously adaptive query processing}. In
  \bibinfo{booktitle}{\emph{Proceedings of the 2000 ACM SIGMOD international
  conference on Management of data}}. \bibinfo{pages}{261--272}.
\newblock


\bibitem[Barber et~al\mbox{.}(2014)]%
        {barber2014memory}
\bibfield{author}{\bibinfo{person}{Ronald Barber}, \bibinfo{person}{Guy
  Lohman}, \bibinfo{person}{Ippokratis Pandis}, \bibinfo{person}{Vijayshankar
  Raman}, \bibinfo{person}{Richard Sidle}, \bibinfo{person}{Gopi Attaluri},
  \bibinfo{person}{Naresh Chainani}, \bibinfo{person}{Sam Lightstone}, {and}
  \bibinfo{person}{David Sharpe}.} \bibinfo{year}{2014}\natexlab{}.
\newblock \showarticletitle{Memory-efficient hash joins}.
\newblock \bibinfo{journal}{\emph{Proceedings of the VLDB Endowment}}
  \bibinfo{volume}{8}, \bibinfo{number}{4} (\bibinfo{year}{2014}),
  \bibinfo{pages}{353--364}.
\newblock


\bibitem[Blanas et~al\mbox{.}(2011)]%
        {blanas2011design}
\bibfield{author}{\bibinfo{person}{Spyros Blanas}, \bibinfo{person}{Yinan Li},
  {and} \bibinfo{person}{Jignesh~M Patel}.} \bibinfo{year}{2011}\natexlab{}.
\newblock \showarticletitle{Design and evaluation of main memory hash join
  algorithms for multi-core CPUs}. In \bibinfo{booktitle}{\emph{Proceedings of
  the 2011 ACM SIGMOD International Conference on Management of data}}.
  \bibinfo{pages}{37--48}.
\newblock


\bibitem[Bloom(1970)]%
        {bloom1970space}
\bibfield{author}{\bibinfo{person}{Burton~H Bloom}.}
  \bibinfo{year}{1970}\natexlab{}.
\newblock \showarticletitle{Space/time trade-offs in hash coding with allowable
  errors}.
\newblock \bibinfo{journal}{\emph{Commun. ACM}} \bibinfo{volume}{13},
  \bibinfo{number}{7} (\bibinfo{year}{1970}), \bibinfo{pages}{422--426}.
\newblock


\bibitem[Boncz et~al\mbox{.}(2005)]%
        {boncz2005monetdb}
\bibfield{author}{\bibinfo{person}{Peter~A Boncz}, \bibinfo{person}{Marcin
  Zukowski}, {and} \bibinfo{person}{Niels Nes}.}
  \bibinfo{year}{2005}\natexlab{}.
\newblock \showarticletitle{MonetDB/X100: Hyper-Pipelining Query Execution.}.
  In \bibinfo{booktitle}{\emph{Cidr}}, Vol.~\bibinfo{volume}{5}.
  \bibinfo{pages}{225--237}.
\newblock


\bibitem[Bratbergsengen(1984)]%
        {bratbergsengen1984hashing}
\bibfield{author}{\bibinfo{person}{Kjell Bratbergsengen}.}
  \bibinfo{year}{1984}\natexlab{}.
\newblock \showarticletitle{Hashing methods and relational algebra operations}.
  In \bibinfo{booktitle}{\emph{Proceedings of the 10th International Conference
  on Very Large Data Bases}}. \bibinfo{pages}{323--333}.
\newblock


\bibitem[Chu et~al\mbox{.}(2015)]%
        {chu2015theory}
\bibfield{author}{\bibinfo{person}{Shumo Chu}, \bibinfo{person}{Magdalena
  Balazinska}, {and} \bibinfo{person}{Dan Suciu}.}
  \bibinfo{year}{2015}\natexlab{}.
\newblock \showarticletitle{From theory to practice: Efficient join query
  evaluation in a parallel database system}. In
  \bibinfo{booktitle}{\emph{Proceedings of the 2015 ACM SIGMOD International
  Conference on Management of Data}}. \bibinfo{pages}{63--78}.
\newblock


\bibitem[Cohen and Matias(2003)]%
        {cohen2003spectral}
\bibfield{author}{\bibinfo{person}{Saar Cohen} {and} \bibinfo{person}{Yossi
  Matias}.} \bibinfo{year}{2003}\natexlab{}.
\newblock \showarticletitle{Spectral bloom filters}. In
  \bibinfo{booktitle}{\emph{Proceedings of the 2003 ACM SIGMOD international
  conference on Management of data}}. \bibinfo{pages}{241--252}.
\newblock


\bibitem[DeWitt et~al\mbox{.}(1993)]%
        {dewitt1993nested}
\bibfield{author}{\bibinfo{person}{David~J DeWitt}, \bibinfo{person}{Jeffrey~F
  Naughton}, {and} \bibinfo{person}{Joseph Burger}.}
  \bibinfo{year}{1993}\natexlab{}.
\newblock \showarticletitle{Nested loops revisited}. In
  \bibinfo{booktitle}{\emph{[1993] Proceedings of the Second International
  Conference on Parallel and Distributed Information Systems}}. IEEE,
  \bibinfo{pages}{230--242}.
\newblock


\bibitem[Ding et~al\mbox{.}(2020)]%
        {ding2020bitvector}
\bibfield{author}{\bibinfo{person}{Bailu Ding}, \bibinfo{person}{Surajit
  Chaudhuri}, {and} \bibinfo{person}{Vivek Narasayya}.}
  \bibinfo{year}{2020}\natexlab{}.
\newblock \showarticletitle{Bitvector-aware query optimization for decision
  support queries}. In \bibinfo{booktitle}{\emph{Proceedings of the 2020 ACM
  SIGMOD International Conference on Management of Data}}.
  \bibinfo{pages}{2011--2026}.
\newblock


\bibitem[Etaati and Etaati(2019)]%
        {etaati2019azure}
\bibfield{author}{\bibinfo{person}{Leila Etaati} {and} \bibinfo{person}{Leila
  Etaati}.} \bibinfo{year}{2019}\natexlab{}.
\newblock \showarticletitle{Azure databricks}.
\newblock \bibinfo{journal}{\emph{Machine Learning with Microsoft Technologies:
  Selecting the Right Architecture and Tools for Your Project}}
  (\bibinfo{year}{2019}), \bibinfo{pages}{159--171}.
\newblock


\bibitem[Freitag et~al\mbox{.}(2020)]%
        {freitag2020adopting}
\bibfield{author}{\bibinfo{person}{Michael Freitag},
  \bibinfo{person}{Maximilian Bandle}, \bibinfo{person}{Tobias Schmidt},
  \bibinfo{person}{Alfons Kemper}, {and} \bibinfo{person}{Thomas Neumann}.}
  \bibinfo{year}{2020}\natexlab{}.
\newblock \showarticletitle{Adopting worst-case optimal joins in relational
  database systems}.
\newblock \bibinfo{journal}{\emph{Proceedings of the VLDB Endowment}}
  \bibinfo{volume}{13}, \bibinfo{number}{12} (\bibinfo{year}{2020}),
  \bibinfo{pages}{1891--1904}.
\newblock


\bibitem[Gehrke et~al\mbox{.}(2003)]%
        {gehrke2003overview}
\bibfield{author}{\bibinfo{person}{Johannes Gehrke}, \bibinfo{person}{Paul
  Ginsparg}, {and} \bibinfo{person}{Jon Kleinberg}.}
  \bibinfo{year}{2003}\natexlab{}.
\newblock \showarticletitle{Overview of the 2003 KDD Cup}.
\newblock \bibinfo{journal}{\emph{Acm Sigkdd Explorations Newsletter}}
  \bibinfo{volume}{5}, \bibinfo{number}{2} (\bibinfo{year}{2003}),
  \bibinfo{pages}{149--151}.
\newblock


\bibitem[Goodman et~al\mbox{.}(1983)]%
        {goodman1983gyo}
\bibfield{author}{\bibinfo{person}{Nathan Goodman}, \bibinfo{person}{Oded
  Shmueli}, {and} \bibinfo{person}{Yong~Chiang Tay}.}
  \bibinfo{year}{1983}\natexlab{}.
\newblock \showarticletitle{GYO reductions, canonical connections, tree and
  cyclic schemas and tree projections}. In
  \bibinfo{booktitle}{\emph{Proceedings of the 2nd ACM SIGACT-SIGMOD symposium
  on Principles of database systems}}. \bibinfo{pages}{267--278}.
\newblock


\bibitem[Graefe et~al\mbox{.}(1998)]%
        {graefe1998hash}
\bibfield{author}{\bibinfo{person}{Goetz Graefe}, \bibinfo{person}{Ross
  Bunker}, {and} \bibinfo{person}{Shaun Cooper}.}
  \bibinfo{year}{1998}\natexlab{}.
\newblock \showarticletitle{Hash joins and hash teams in Microsoft SQL Server}.
  In \bibinfo{booktitle}{\emph{VLDB}}, Vol.~\bibinfo{volume}{98}.
  \bibinfo{pages}{86--97}.
\newblock


\bibitem[Idris et~al\mbox{.}(2017)]%
        {idris2017dynamic}
\bibfield{author}{\bibinfo{person}{Muhammad Idris},
  \bibinfo{person}{Mart{\'\i}n Ugarte}, {and} \bibinfo{person}{Stijn
  Vansummeren}.} \bibinfo{year}{2017}\natexlab{}.
\newblock \showarticletitle{The dynamic yannakakis algorithm: Compact and
  efficient query processing under updates}. In
  \bibinfo{booktitle}{\emph{Proceedings of the 2017 ACM International
  Conference on Management of Data}}. \bibinfo{pages}{1259--1274}.
\newblock


\bibitem[Kemper et~al\mbox{.}(1999)]%
        {kemper1999generalised}
\bibfield{author}{\bibinfo{person}{Alfons Kemper}, \bibinfo{person}{Donald
  Kossmann}, {and} \bibinfo{person}{Christian Wiesner}.}
  \bibinfo{year}{1999}\natexlab{}.
\newblock \showarticletitle{Generalised hash teams for join and group-by}. In
  \bibinfo{booktitle}{\emph{VLDB}}, Vol.~\bibinfo{volume}{99}. Citeseer,
  \bibinfo{pages}{30--41}.
\newblock


\bibitem[Kirsch and Mitzenmacher(2006)]%
        {kirsch2006less}
\bibfield{author}{\bibinfo{person}{Adam Kirsch} {and} \bibinfo{person}{Michael
  Mitzenmacher}.} \bibinfo{year}{2006}\natexlab{}.
\newblock \showarticletitle{Less hashing, same performance: Building a better
  bloom filter}. In \bibinfo{booktitle}{\emph{Algorithms--ESA 2006: 14th Annual
  European Symposium, Zurich, Switzerland, September 11-13, 2006. Proceedings
  14}}. Springer, \bibinfo{pages}{456--467}.
\newblock


\bibitem[Koutris(2011)]%
        {koutris2011bloom}
\bibfield{author}{\bibinfo{person}{Paraschos Koutris}.}
  \bibinfo{year}{2011}\natexlab{}.
\newblock \showarticletitle{Bloom filters in distributed query execution}.
\newblock \bibinfo{journal}{\emph{University of Washington, CSE}}
  \bibinfo{volume}{544} (\bibinfo{year}{2011}).
\newblock


\bibitem[Lahiri et~al\mbox{.}(2013)]%
        {lahiri2013oracle}
\bibfield{author}{\bibinfo{person}{Tirthankar Lahiri},
  \bibinfo{person}{Marie-Anne Neimat}, {and} \bibinfo{person}{Steve Folkman}.}
  \bibinfo{year}{2013}\natexlab{}.
\newblock \showarticletitle{Oracle TimesTen: An In-Memory Database for
  Enterprise Applications.}
\newblock \bibinfo{journal}{\emph{IEEE Data Eng. Bull.}} \bibinfo{volume}{36},
  \bibinfo{number}{2} (\bibinfo{year}{2013}), \bibinfo{pages}{6--13}.
\newblock


\bibitem[Leskovec et~al\mbox{.}(2010a)]%
        {leskovec2010predicting}
\bibfield{author}{\bibinfo{person}{Jure Leskovec}, \bibinfo{person}{Daniel
  Huttenlocher}, {and} \bibinfo{person}{Jon Kleinberg}.}
  \bibinfo{year}{2010}\natexlab{a}.
\newblock \showarticletitle{Predicting positive and negative links in online
  social networks}. In \bibinfo{booktitle}{\emph{Proceedings of the 19th
  international conference on World wide web}}. \bibinfo{pages}{641--650}.
\newblock


\bibitem[Leskovec et~al\mbox{.}(2010b)]%
        {leskovec2010signed}
\bibfield{author}{\bibinfo{person}{Jure Leskovec}, \bibinfo{person}{Daniel
  Huttenlocher}, {and} \bibinfo{person}{Jon Kleinberg}.}
  \bibinfo{year}{2010}\natexlab{b}.
\newblock \showarticletitle{Signed networks in social media}. In
  \bibinfo{booktitle}{\emph{Proceedings of the SIGCHI conference on human
  factors in computing systems}}. \bibinfo{pages}{1361--1370}.
\newblock


\bibitem[Leskovec et~al\mbox{.}(2005)]%
        {leskovec2005graphs}
\bibfield{author}{\bibinfo{person}{Jure Leskovec}, \bibinfo{person}{Jon
  Kleinberg}, {and} \bibinfo{person}{Christos Faloutsos}.}
  \bibinfo{year}{2005}\natexlab{}.
\newblock \showarticletitle{Graphs over time: densification laws, shrinking
  diameters and possible explanations}. In
  \bibinfo{booktitle}{\emph{Proceedings of the eleventh ACM SIGKDD
  international conference on Knowledge discovery in data mining}}.
  \bibinfo{pages}{177--187}.
\newblock


\bibitem[Lo and Ravishankar(1996)]%
        {lo1996spatial}
\bibfield{author}{\bibinfo{person}{Ming-Ling Lo} {and}
  \bibinfo{person}{Chinya~V Ravishankar}.} \bibinfo{year}{1996}\natexlab{}.
\newblock \showarticletitle{Spatial hash-joins}. In
  \bibinfo{booktitle}{\emph{Proceedings of the 1996 ACM SIGMOD international
  conference on Management of data}}. \bibinfo{pages}{247--258}.
\newblock


\bibitem[Luo et~al\mbox{.}(2018)]%
        {luo2018optimizing}
\bibfield{author}{\bibinfo{person}{Lailong Luo}, \bibinfo{person}{Deke Guo},
  \bibinfo{person}{Richard~TB Ma}, \bibinfo{person}{Ori Rottenstreich}, {and}
  \bibinfo{person}{Xueshan Luo}.} \bibinfo{year}{2018}\natexlab{}.
\newblock \showarticletitle{Optimizing bloom filter: Challenges, solutions, and
  comparisons}.
\newblock \bibinfo{journal}{\emph{IEEE Communications Surveys \& Tutorials}}
  \bibinfo{volume}{21}, \bibinfo{number}{2} (\bibinfo{year}{2018}),
  \bibinfo{pages}{1912--1949}.
\newblock


\bibitem[Mackert and Lohman(1986)]%
        {mackert1986r}
\bibfield{author}{\bibinfo{person}{Lothar~F Mackert} {and}
  \bibinfo{person}{Guy~M Lohman}.} \bibinfo{year}{1986}\natexlab{}.
\newblock \showarticletitle{R* optimizer validation and performance evaluation
  for local queries}. In \bibinfo{booktitle}{\emph{Proceedings of the 1986 ACM
  SIGMOD international conference on Management of data}}.
  \bibinfo{pages}{84--95}.
\newblock


\bibitem[Michael et~al\mbox{.}(2007)]%
        {michael2007improving}
\bibfield{author}{\bibinfo{person}{Loizos Michael}, \bibinfo{person}{Wolfgang
  Nejdl}, \bibinfo{person}{Odysseas Papapetrou}, {and} \bibinfo{person}{Wolf
  Siberski}.} \bibinfo{year}{2007}\natexlab{}.
\newblock \showarticletitle{Improving distributed join efficiency with extended
  bloom filter operations}. In \bibinfo{booktitle}{\emph{21st International
  Conference on Advanced Information Networking and Applications (AINA'07)}}.
  IEEE, \bibinfo{pages}{187--194}.
\newblock


\bibitem[Mullin(1990)]%
        {mullin1990optimal}
\bibfield{author}{\bibinfo{person}{James~K. Mullin}.}
  \bibinfo{year}{1990}\natexlab{}.
\newblock \showarticletitle{Optimal semijoins for distributed database
  systems}.
\newblock \bibinfo{journal}{\emph{IEEE Transactions on Software Engineering}}
  \bibinfo{volume}{16}, \bibinfo{number}{5} (\bibinfo{year}{1990}),
  \bibinfo{pages}{558--560}.
\newblock


\bibitem[Nes and Kersten(2012)]%
        {nes2012monetdb}
\bibfield{author}{\bibinfo{person}{Stratos Idreos Fabian Groffen~Niels Nes}
  {and} \bibinfo{person}{Stefan Manegold Sjoerd Mullender~Martin Kersten}.}
  \bibinfo{year}{2012}\natexlab{}.
\newblock \showarticletitle{MonetDB: Two decades of research in column-oriented
  database architectures}.
\newblock \bibinfo{journal}{\emph{Data Engineering}}  \bibinfo{volume}{40}
  (\bibinfo{year}{2012}).
\newblock


\bibitem[Ngo et~al\mbox{.}(2014a)]%
        {ngo2014beyond}
\bibfield{author}{\bibinfo{person}{Hung~Q Ngo}, \bibinfo{person}{Dung~T
  Nguyen}, \bibinfo{person}{Christopher Re}, {and} \bibinfo{person}{Atri
  Rudra}.} \bibinfo{year}{2014}\natexlab{a}.
\newblock \showarticletitle{Beyond worst-case analysis for joins with
  minesweeper}. In \bibinfo{booktitle}{\emph{Proceedings of the 33rd ACM
  SIGMOD-SIGACT-SIGART symposium on Principles of database systems}}.
  \bibinfo{pages}{234--245}.
\newblock


\bibitem[Ngo et~al\mbox{.}(2018)]%
        {ngo2018worst}
\bibfield{author}{\bibinfo{person}{Hung~Q Ngo}, \bibinfo{person}{Ely Porat},
  \bibinfo{person}{Christopher R{\'e}}, {and} \bibinfo{person}{Atri Rudra}.}
  \bibinfo{year}{2018}\natexlab{}.
\newblock \showarticletitle{Worst-case optimal join algorithms}.
\newblock \bibinfo{journal}{\emph{Journal of the ACM (JACM)}}
  \bibinfo{volume}{65}, \bibinfo{number}{3} (\bibinfo{year}{2018}),
  \bibinfo{pages}{1--40}.
\newblock


\bibitem[Ngo et~al\mbox{.}(2014b)]%
        {ngo2014skew}
\bibfield{author}{\bibinfo{person}{Hung~Q Ngo}, \bibinfo{person}{Christopher
  R{\'e}}, {and} \bibinfo{person}{Atri Rudra}.}
  \bibinfo{year}{2014}\natexlab{b}.
\newblock \showarticletitle{Skew strikes back: New developments in the theory
  of join algorithms}.
\newblock \bibinfo{journal}{\emph{Acm Sigmod Record}} \bibinfo{volume}{42},
  \bibinfo{number}{4} (\bibinfo{year}{2014}), \bibinfo{pages}{5--16}.
\newblock


\bibitem[Patel and DeWitt(1996)]%
        {patel1996partition}
\bibfield{author}{\bibinfo{person}{Jignesh~M Patel} {and}
  \bibinfo{person}{David~J DeWitt}.} \bibinfo{year}{1996}\natexlab{}.
\newblock \showarticletitle{Partition based spatial-merge join}.
\newblock \bibinfo{journal}{\emph{ACM Sigmod Record}} \bibinfo{volume}{25},
  \bibinfo{number}{2} (\bibinfo{year}{1996}), \bibinfo{pages}{259--270}.
\newblock


\bibitem[Poess and Floyd(2000)]%
        {poess2000new}
\bibfield{author}{\bibinfo{person}{Meikel Poess} {and} \bibinfo{person}{Chris
  Floyd}.} \bibinfo{year}{2000}\natexlab{}.
\newblock \showarticletitle{New TPC benchmarks for decision support and web
  commerce}.
\newblock \bibinfo{journal}{\emph{ACM Sigmod Record}} \bibinfo{volume}{29},
  \bibinfo{number}{4} (\bibinfo{year}{2000}), \bibinfo{pages}{64--71}.
\newblock


\bibitem[R{\u{a}}ducanu et~al\mbox{.}(2013)]%
        {ruaducanu2013micro}
\bibfield{author}{\bibinfo{person}{Bogdan R{\u{a}}ducanu},
  \bibinfo{person}{Peter Boncz}, {and} \bibinfo{person}{Marcin Zukowski}.}
  \bibinfo{year}{2013}\natexlab{}.
\newblock \showarticletitle{Micro adaptivity in vectorwise}. In
  \bibinfo{booktitle}{\emph{Proceedings of the 2013 ACM SIGMOD International
  Conference on Management of Data}}. \bibinfo{pages}{1231--1242}.
\newblock


\bibitem[Severance and Lohman(1976)]%
        {severance1976differential}
\bibfield{author}{\bibinfo{person}{Dennis~G Severance} {and}
  \bibinfo{person}{Guy~M Lohman}.} \bibinfo{year}{1976}\natexlab{}.
\newblock \showarticletitle{Differential files: Their application to the
  maintenance of large databases}.
\newblock \bibinfo{journal}{\emph{ACM Transactions on Database Systems (TODS)}}
  \bibinfo{volume}{1}, \bibinfo{number}{3} (\bibinfo{year}{1976}),
  \bibinfo{pages}{256--267}.
\newblock


\bibitem[Shanghooshabad and Triantafillou(2022)]%
        {shanghooshabad2022graphical}
\bibfield{author}{\bibinfo{person}{Ali~Mohammadi Shanghooshabad} {and}
  \bibinfo{person}{Peter Triantafillou}.} \bibinfo{year}{2022}\natexlab{}.
\newblock \showarticletitle{Graphical Join: A New Physical Join Algorithm for
  RDBMSs}.
\newblock \bibinfo{journal}{\emph{arXiv preprint arXiv:2206.10435}}
  (\bibinfo{year}{2022}).
\newblock


\bibitem[Stonebraker and Rowe(1986)]%
        {stonebraker1986design}
\bibfield{author}{\bibinfo{person}{Michael Stonebraker} {and}
  \bibinfo{person}{Lawrence~A Rowe}.} \bibinfo{year}{1986}\natexlab{}.
\newblock \showarticletitle{The design of Postgres}.
\newblock \bibinfo{journal}{\emph{ACM Sigmod Record}} \bibinfo{volume}{15},
  \bibinfo{number}{2} (\bibinfo{year}{1986}), \bibinfo{pages}{340--355}.
\newblock


\bibitem[Veldhuizen(2014)]%
        {veldhuizen2014leapfrog}
\bibfield{author}{\bibinfo{person}{Todd~L Veldhuizen}.}
  \bibinfo{year}{2014}\natexlab{}.
\newblock \showarticletitle{Leapfrog triejoin: A simple, worst-case optimal
  join algorithm}. In \bibinfo{booktitle}{\emph{Proc. International Conference
  on Database Theory}}.
\newblock


\bibitem[Wu et~al\mbox{.}(2014)]%
        {wu2014multipredicate}
\bibfield{author}{\bibinfo{person}{Haicheng Wu}, \bibinfo{person}{Daniel Zinn},
  \bibinfo{person}{Molham Aref}, {and} \bibinfo{person}{Sudhakar
  Yalamanchili}.} \bibinfo{year}{2014}\natexlab{}.
\newblock \showarticletitle{Multipredicate join algorithms for accelerating
  relational graph processing on GPUs}. In
  \bibinfo{booktitle}{\emph{International Workshop on Accelerating Data
  Management Systems Using Modern Processor and Storage Architectures}},
  Vol.~\bibinfo{volume}{10}.
\newblock


\bibitem[Yannakakis(1981)]%
        {yannakakis1981algorithms}
\bibfield{author}{\bibinfo{person}{Mihalis Yannakakis}.}
  \bibinfo{year}{1981}\natexlab{}.
\newblock \showarticletitle{Algorithms for acyclic database schemes}. In
  \bibinfo{booktitle}{\emph{VLDB}}, Vol.~\bibinfo{volume}{81}.
  \bibinfo{pages}{82--94}.
\newblock


\bibitem[Zhao et~al\mbox{.}(2018)]%
        {zhao2018random}
\bibfield{author}{\bibinfo{person}{Zhuoyue Zhao}, \bibinfo{person}{Robert
  Christensen}, \bibinfo{person}{Feifei Li}, \bibinfo{person}{Xiao Hu}, {and}
  \bibinfo{person}{Ke Yi}.} \bibinfo{year}{2018}\natexlab{}.
\newblock \showarticletitle{Random sampling over joins revisited}. In
  \bibinfo{booktitle}{\emph{Proceedings of the 2018 International Conference on
  Management of Data}}. \bibinfo{pages}{1525--1539}.
\newblock


\bibitem[Zhu et~al\mbox{.}(2017)]%
        {zhu2017looking}
\bibfield{author}{\bibinfo{person}{Jianqiao Zhu}, \bibinfo{person}{Navneet
  Potti}, \bibinfo{person}{Saket Saurabh}, {and} \bibinfo{person}{Jignesh~M
  Patel}.} \bibinfo{year}{2017}\natexlab{}.
\newblock \showarticletitle{Looking ahead makes query plans robust: Making the
  initial case with in-memory star schema data warehouse workloads}.
\newblock \bibinfo{journal}{\emph{Proceedings of the VLDB Endowment}}
  \bibinfo{volume}{10}, \bibinfo{number}{8} (\bibinfo{year}{2017}),
  \bibinfo{pages}{889--900}.
\newblock


\bibitem[Zou et~al\mbox{.}(2011)]%
        {zou2011gstore}
\bibfield{author}{\bibinfo{person}{Lei Zou}, \bibinfo{person}{Jinghui Mo},
  \bibinfo{person}{Lei Chen}, \bibinfo{person}{M~Tamer {\"O}zsu}, {and}
  \bibinfo{person}{Dongyan Zhao}.} \bibinfo{year}{2011}\natexlab{}.
\newblock \showarticletitle{gStore: answering SPARQL queries via subgraph
  matching}.
\newblock \bibinfo{journal}{\emph{Proceedings of the VLDB Endowment}}
  \bibinfo{volume}{4}, \bibinfo{number}{8} (\bibinfo{year}{2011}),
  \bibinfo{pages}{482--493}.
\newblock


\bibitem[Zukowski et~al\mbox{.}(2012)]%
        {zukowski2012vectorwise}
\bibfield{author}{\bibinfo{person}{Marcin Zukowski}, \bibinfo{person}{Mark
  Van~de Wiel}, {and} \bibinfo{person}{Peter Boncz}.}
  \bibinfo{year}{2012}\natexlab{}.
\newblock \showarticletitle{Vectorwise: A vectorized analytical DBMS}. In
  \bibinfo{booktitle}{\emph{2012 IEEE 28th International Conference on Data
  Engineering}}. IEEE, \bibinfo{pages}{1349--1350}.
\newblock


\end{thebibliography}

\end{document}